# The spherically symmetric collapse of a massless scalar field

**Rufus S Hamadé and John M Stewart**

Department of Applied Mathematics and Theoretical Physics Silver Street, Cambridge CB3 9EW, UK.

**Abstract.** We report on a numerical study of the spherically symmetric collapse of a self-gravitating massless scalar field. Earlier results of Choptuik(1992, 1994) are confirmed. The field either disperses to infinity or collapses to a black hole, depending on the strength of the initial data. For evolutions where the strength is close to but below the strength required to form a black hole, we argue that there will be a region close to the axis where the scalar curvature and field energy density can reach arbitrarily large levels, and which is visible to distant observers

## 1. Introduction

This paper reports a numerical study of the Einstein equations

$$R_{\mu\nu} - \tfrac{1}{2} R g_{\mu\nu} = 8\pi G T_{\mu\nu}, \tag{1.1}$$

where the energy-momentum tensor is that of a real massless scalar field

$$T_{\mu\nu} = \Psi_{,\mu}\Psi_{,\nu} - \tfrac{1}{2} g_{\mu\nu} g^{\alpha\beta}\Psi_{,\alpha}\Psi_{,\beta}. \tag{1.2}$$

Equations (1.1–2) imply

$$R_{\mu\nu} = 8\pi G \Psi_{,\mu}\Psi_{,\nu}, \tag{1.3}$$

and energy conservation $T^{\mu\nu}{}_{;\nu} = 0$ (or the twice contracted Bianchi identities) imply

$$\Box\Psi \equiv \Psi^{,\mu}{}_{;\mu} = 0. \tag{1.4}$$

Attention is restricted to a spherically symmetric geometry and field $\Psi$.

This problem has been studied analytically in a series of papers by Christodoulou, of which Christodoulou(1991, 1993) are particularly relevant here. For initial data in a neighbourhood of trivial data, one might expect the field to disperse to future null infinity leaving behind a flat spacetime. However stronger initial data might be expected to produce a black hole.

This idea was taken up by Choptuik(1992, 1994) who proposed the following programme. Set up a well-posed initial value problem for the system governed by equations (1.3–4). Choose consistent initial data dependent on a parameter $p$ such that small $p$ represents near trivial data, while large $p$ leads to black hole formation. By continuity there will be a *critical value* $p^*$ of the parameter such that black holes only occur for $p \geq p^*$. What do near critical solutions (those given by $p \approx p^*$) look like? Choptuik chose an arbitrary time variable $t$ and area radius $r$ as his principal coordinates, setting

$$ds^2 = -\alpha^2(t,r)\,dt^2 + a^2(t,r)\,dr^2 + r^2 d\Omega^2, \tag{1.5}$$

where $d\Omega^2 = d\theta^2 + \sin^2\theta\, d\phi^2$ is the metric of the unit sphere. The area coordinate $r$ has a physical meaning as does

$$T = \int_0^t \alpha(w,0)\,dw, \tag{1.6}$$

the proper time on axis. He presented numerical evidence that in the $p \to p^*$ limit the strong field evolution is *universal*, i.e., the same for all choices of initial data, and generates structure on arbitrarily small spacetime scales. Further he argued that for $p > p^*$ the resultant black hole has a mass $m_{BH} \propto (p - p^*)^\gamma$, where $\gamma \approx 0.375$ is a universal exponent.

Some attempts, Gundlach *et al* (1994), Hübner(1994) and Garfinkle(1994), have been made to reproduce Choptuik's results, but there is a difficulty. Because structure appears on ever finer scales, accuracy will only be maintained if the spacetime grid used to simulate numerically the field equations is refined to corresponding levels. Simple algorithms use fixed grid spacings and for this problem $\sim 10^{20}$ grid points would be needed; this is far more than the largest computer could handle. Choptuik made use of an adaptive mesh refinement algorithm proposed by Berger & Oliger(1984). This creates the necessary fine grids only where and when they are needed, and destroys them when they are no longer useful. Such algorithms are exceedingly complicated to write and implement, but without them a study of this topic would seem to be intractable.

We have written two versions of the Berger & Oliger algorithm in order to study the Choptuik problem. We wanted to study in particular the strong-field regions. A disadvantage of Choptuik's coordinate system (and indeed of any chart which uses an area coordinate $r$) is that it breaks down at an apparent horizon. We therefore decided to use one or more null coordinates and to set up a characteristic initial value problem. Our initial algorithm used $(u, w, \theta, \phi)$ coordinates. Here $u$ is retarded time, and $w$ is an affine parameter along the generators of the future, outgoing light cones $u =$ const. Following Christodoulou(1991) we took the line element to be

$$ds^2 = -a(u,w)\,du^2 - 2\,du\,dw + r(u,w)\,d\Omega^2. \tag{1.7}$$

This is an excellent choice when the the evolution is far from critical, but the algorithm becomes numerically unstable for near-critical evolutions. (The function $a(u, w)$ is effectively a phase velocity for the field, which both oscillates in sign and grows in magnitude, thereby violating the CFL condition for any explicit algorithm.) Following Christodoulou(1993) we switched to double null coordinates $(u, v, \theta, \phi)$, where $v$ is advanced time.

Considerable care is needed to obtain a numerically stable algorithm, and the appropriate choice of variables and equations is reported in section 2. The salient features of our code are discussed in section 3, and its results are adumbrated in section 4. Section 5 details our conclusions. We confirm Choptuik's results, and in addition argue that for subcritical evolutions the strong field regions are visible to distant observers. For the impatient reader we have tried to make the figure captions as self-contained as possible.

## 2. Coordinates, variables and equations

Spacetime is spherically symmetric $\mathcal{M} = \mathcal{N} \times \mathcal{S}^2$ where $\mathcal{N}$ is a 2-dimensional pseudo-Riemannian manifold of signature $(-+)$, and $\mathcal{S}^2$ has the topology of the unit sphere. Let $(\theta, \varphi)$ be the standard polar coordinates on the unit sphere. Let $\Gamma$ denote the central worldline in $\mathcal{N}$. Choose an arbitrary parameter $t$ along $\Gamma$ increasing into the future. Denote the future/past null cones of points on $\Gamma$ by $C_+(t)/C_-(t)$ respectively. In a regular spacetime an arbitrary point of $\mathcal{N}$ lies on precisely one cone of each family, say $C_+(u)$, $C_-(v)$. Then $(u, v)$ is a coordinate chart in $\mathcal{N}$ and $(u, v, \theta, \varphi)$ is a coordinate chart in $\mathcal{M}$. Note that this fixes the axis $\Gamma$ to be $u = v$. The remaining freedom in $(u, v)$ is the freedom of the choice of $t$ on $\Gamma$, or equivalently the freedom of the choice of $v$ on some initial future null cone, say $C_+(0)$.

The line element may be taken as, c.f., Christodoulou(1993),

$$ds^2 = -a^2(u,v)\, du\, dv + r^2(u,v)\, d\Omega^2. \tag{2.1}$$

Now that we have fixed the coordinate chart, we may adopt the usual convention of the partial differential equation literature where, e.g., $\partial f(u,v)/\partial u$ is abbreviated to $f_u$. Further we label equations, as indicated below. The scalar field equation (1.4) becomes

$$E_0 \equiv r\Psi_{uv} + r_u\Psi_v + r_v\Psi_u = 0. \tag{2.2}$$

The Einstein field equations (1.3) expand to

$$\begin{aligned}
E1 &\equiv r r_{uv} + r_u r_v + \tfrac{1}{4}a^2 = 0,\\
E2 &\equiv a^{-1}a_{uv} - a^{-2}a_u a_v + r^{-1}r_{uv} + 4\pi G\Psi_u\Psi_v = 0,\\
E3 &\equiv r_{uu} - 2a^{-1}a_u r_u + 4\pi G r\Psi_u^2 = 0,\\
E4 &\equiv r_{vv} - 2a^{-1}a_v r_v + 4\pi G r\Psi_v^2 = 0.
\end{aligned} \tag{2.3}$$

It is tempting to try to construct a numerical scheme to solve a subset of these equations. (They are not independent.) Such a scheme would have the merits of simplicity and conciseness, but we have been unable to derive a stable explicit scheme for the equation $E2$. We have therefore reformulated the equations as a first order system.

It is convenient to introduce new variables

$$c = \frac{a_u}{a}, \quad d = \frac{a_v}{a}, \quad f = r_u, \quad g = r_v, \quad s = \sqrt{4\pi G}\Psi, \quad p = s_u, \quad q = s_v, \tag{2.4}$$

as well as the auxiliary quantities

$$\lambda = fg + \tfrac{1}{4}a^2, \qquad \mu = fq + gp. \tag{2.5}$$

The system of equations (2.1–3) now expands to

$$
\begin{aligned}
F1 &\equiv a_u - ac = 0,\\
F2 &\equiv a_v - ad = 0,\\
F3 &\equiv r_u - f = 0,\\
F4 &\equiv r_v - g = 0,\\
F5 &\equiv s_u - p = 0,\\
F6 &\equiv s_v - q = 0,\\
F7 &\equiv rp_v + \mu = 0,\\
F8 &\equiv rq_u + \mu = 0,\\
F9 &\equiv r^2 c_v - \lambda + r^2 pq = 0,\\
F10 &\equiv r^2 d_u - \lambda + r^2 pq = 0,\\
F11 &\equiv rf_v + \lambda = 0,\\
F12 &\equiv rg_u + \lambda = 0,\\
F13 &\equiv f_u - 2cf + rp^2 = 0,\\
F14 &\equiv g_v - 2dg + rq^2 = 0.
\end{aligned}
\tag{2.6}
$$

Of course not all of these equations are independent, and indeed some of the dependent variables may well be irrelevant. Scalar quantities of physical interest include the Hawking mass

$$
m(u,v) = \frac{r}{2}\left(1 + \frac{4fg}{a^2}\right), \tag{2.7}
$$

the Ricci scalar curvature

$$
R(u,v) = -\frac{8pq}{a^2}, \tag{2.8}
$$

and the energy density $\rho$ of the scalar field. This may be defined as follows. Suppose we introduce time $t$ and radial $r^*$ coordinates via $u = t - r^*$, $v = t + r^*$. Then

$$
\rho(u,v) = T_t{}^t = \frac{1}{8\pi G a^2}\left(p^2 + q^2\right), \tag{2.9}
$$

where the energy-momentum tensor was defined by equation (1.2). The vanishing of the quantity $g$ signifies the apparent horizon. Another quantity of physical interest is the proper time $T(u,r)$ as measured by an observer who moves along a worldline of constant $r$. We may invert the relation $r = r(u,v)$ to obtain $v = v(u,r)$ from which $dv = (\partial v/\partial u)_r \, du$ and the line element (2.1) implies

$$
T(u,r) = \int_0^u a(w, v(w,r))\sqrt{\left(\frac{\partial v}{\partial u}\right)_r}\, dw. \tag{2.10}
$$

This simplifies considerably on the axis, $v = u$, to formula (1.6)

$$T(u,0) = \int_0^u a(w,w)\,dw. \tag{2.11}$$

The quantity $c$ does not appear in any of the formulae in the paragraph above, and so may be ignored. Thus we may immediately discard the equations $F1$ and $F9$. Notice next that there is only one equation governing the evolution of each of the quantities $d$ and $q$, viz., $F10$ and $F8$ respectively. We choose to evolve the remaining quantities $a$, $r$, $s$, $p$, $f$ and $g$ by the equations governing their $v$-derivatives, viz., $F2$, $F4$, $F6$, $F7$, $F11$ and $F14$ respectively. The remaining equations are not used directly in the evolution, although they help to determine the boundary conditions.

These equations have to be supplemented by the specification of initial and boundary data. On the initial cone $C_+(0)$ we have to specify $d(0,v)$ and $q(0,v)$ and these data are unconstrained. In practice we assume a flat geometry initially, i.e., $d(0,v) = 0$, which is equivalent to using up the freedom in the $v$-coordinate. On the axis we have to specify values for $a$, $r$, $s$, $p$, $f$ and $g$. Obviously $r = 0$ there and this demands $f = -g$ there. The equation $F11$ requires $\lambda = 0$ on axis and so the first equation (2.5) forces $g = \frac{1}{2}a = -f$ on axis. Equation $F8$ and the second equation (2.5) now ensure that $p = q$ on axis. The boundary values for $a$ and $s$ are obtained by requiring

$$\frac{\partial a}{\partial r} = \frac{\partial s}{\partial r} = 0 \text{ on axis,} \tag{2.12}$$

as will be explained in the next section.

Our evolution scheme can be outlined as follows. On $C_+(0)$ we choose $d = 0$ and $q(0,v)$ arbitrarily. We now integrate outwards from the axis $F2$ to obtain $a$, $F4$ and $F14$ to obtain $r$ and $g$, $F6$ to obtain $s$, $F11$ to find $f$ and $F7$ to obtain $p$. Now that we have a complete solution on $C_+(0)$ we use $F8$ and $F10$ to predict values of $q$ and $d$ on a cone $C_+(h)$ to the future of the initial one, and a similar radial integration completes the solution. This procedure can be repeated until a singularity occurs. The details are explained in the next section.

## 3. The numerical code

Our basic code is extremely simple. We construct a grid in $uv$-space with a spacing of $h$ in both the $u$- and $v$-directions. We write the equations schematically as

$$y_u = F(y,z), \qquad z_v = G(y,z), \tag{3.1}$$

where $y$ denotes variables evolved in the $u$-direction, and $z$ denotes those evolved in the $v$-direction. The basic computation cell is depicted below in figure 1.

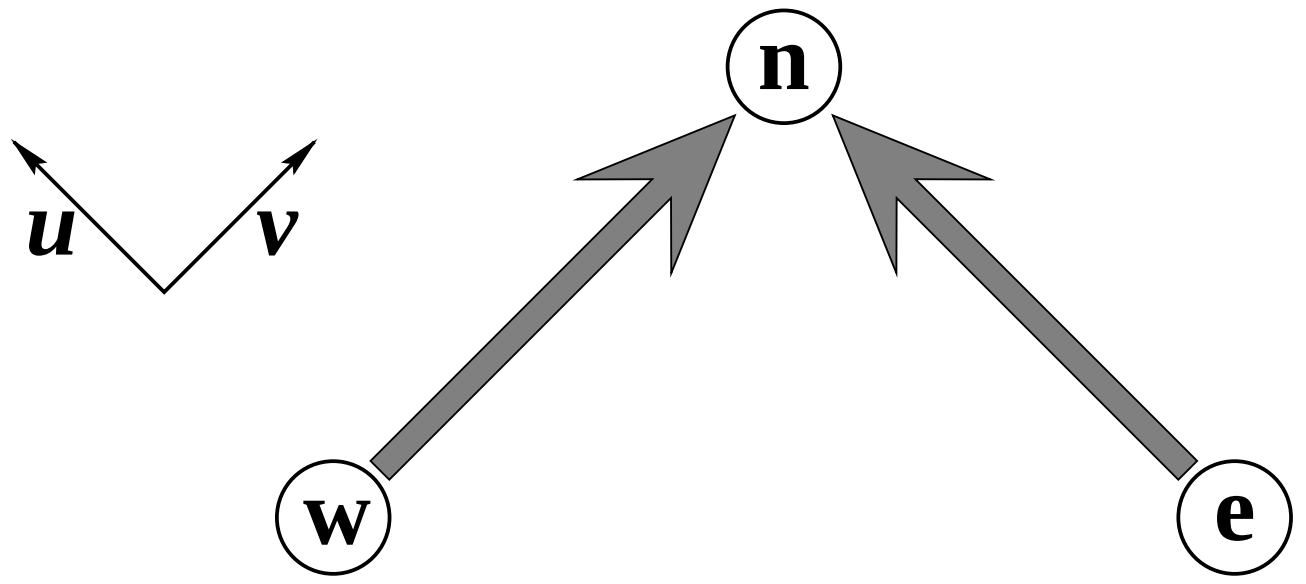

**Fig. 1** The basic computational cell. Knowing the solution at `w`, $(u, v-h)$ and `e`, $(u-h, v)$ allows us to estimate the solution at `n`, $(u, v)$.

We assume that the solution is known at the points `e`, $(u-h, v)$ and `w`, $(u, v-h)$ and we want to predict it at `n`, $(u, v)$. We first predict the $y$-values at `n` via the explicit step

$$\widehat{y}_n = y_e + hF(y_e, z_e). \tag{3.2}$$

Next we predict the $z$-values via the step

$$\widehat{z}_n = z_w + \tfrac{1}{2}h\left(G(y_w, z_w) + G(\widehat{y}_n, \widehat{z}_n)\right). \tag{3.3}$$

Although this appears to be an implicit method the equations can be ordered so as to be linear in the unknown quantity, and the step can be made explicit. For the second stage we correct the prediction via

$$y_n = \tfrac{1}{2}\left(\widehat{y}_n + y_e + hF(\widehat{y}_n, \widehat{z}_n)\right), \tag{3.4}$$

and

$$z_n = \tfrac{1}{2}\left(\widehat{z}_n + z_w + hG(\widehat{y}_n, \widehat{z}_n)\right). \tag{3.5}$$

Although it would have been easier to use the simpler algorithms (3.2) and (3.4) in the $v$-direction, the more complicated (3.3) and (3.5) ensure stability. The completely explicit procedure would involve a Courant number of unity, which is only marginally stable.

All of the boundary conditions were straightforward to implement except those for $a$ and $s$, equation (2.12). The quantity $\partial a/\partial r$ at $(u, u)$ can be approximated using values of $a$ at $(u, u)$, $(u-h, u+h)$ and $(u-2h, u+2h)$, and its vanishing allows us to predict $a(u, u)$. The quantity $s$ is treated similarly. This means that to construct quantities at retarded time $u$ we need to have stored values at times $u-h$ and $u-2h$.

A crucial feature of our code is the adaptive mesh refinement. This is described in detail by Berger and Oliger (1984), but it is appropriate to review it here, since we were able to construct a perfectly satisfactory simplified version. Our simplification uses a doubly linked list of grids, figure 2, rather than the complicated tree of the original version. The original grid is the one we start with. (Typical ranges are $0 \le u \le v \le 2.0$ with a grid spacing of $h = 10^{-3}$.) Each child grid has a grid spacing equal to a quarter of that of its parent. There are two operations which take place on this list, regridding and evolution.

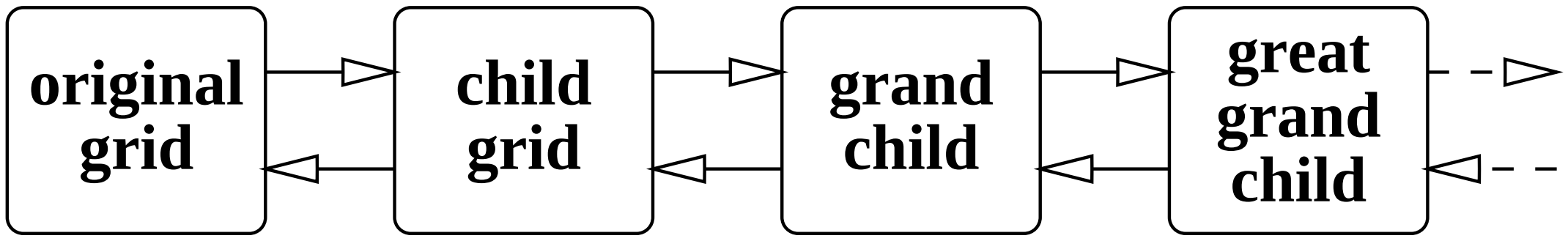

**Fig. 2** The linked list structure for the simplified Berger and Oliger(1984) algorithm. Each grid has equal $u$- and $v$-spacing four times finer than its parent and four times coarser than its child.

After each grid $\mathcal{G}$ has been evolved a fixed number of steps in the $u$-direction since the last revision, it is revised.

(1) We first estimate the local truncation error. Since we store values at three levels of $u$ we can carry out a single evolution step with twice the grid spacing from $u - 2h$ to $u$, which can be compared with the more accurate evolution from $u - h$ to $u$. All points at which the error is unsatisfactory are marked. They may be expected to occur in clusters and buffer zones of satisfactory points are added to the boundaries of each cluster. In our simplified version only one cluster is allowed.

(2) We repeat step (1) for each child of the grid $\mathcal{G}$.

(3) Starting at the finest grid and working back to $\mathcal{G}$ we build new child grids to cover the clusters, interpolating values from their parents. If there are three or more generations we have to ensure that the $v$-range of each grandchild grid is included in the corresponding child range.

(4) If all the child grids of $\mathcal{G}$ are free of error they may be destroyed and the used memory is returned to the computer's operating system.

The regridding operation creates new subgrids where and when they are needed, and removes them when they are obsolescent. We move down the doubly linked list making error estimates, and back up the list to $\mathcal{G}$ when rebuilding them.

The evolution operation also involves the linked list.

(5) Each grid $\mathcal{G}$ is evolved one step in the $u$-direction.

(6) Proceeding down the list from $\mathcal{G}$ each child grid is evolved to the $u$-level of its parent. If boundary data are needed this is obtained by interpolation from the parent grid.

(7) Once all child grids have reached the $u$-level of the parent, injection occurs. Starting at the finest grid accurate values are injected from each child into the parent grid.

Thus evolution starts with the coarsest grid and ends with the finest, while the injection proceeds in the opposite direction. After each injection step a grid contains the most accurate data available.

Injection proved to be a source of noise. Once a grid has had accurate values injected over a $v$-range $[v_1, v_2]$ we found occasionally some discontinuities at $v = v_2$. Ignoring them led to the introduction of high frequency noise which corrupted our results. A simple, crude fix would have been to introduce some explicit artificial viscosity to damp out the noise. However we found it to be more convenient, and also more accurate, to redo the $v$-integration for $v > v_2$ after each injection.

The adaptive mesh refinement algorithm is most simply implemented using recursion and our codes were written in the programming language C. We have implemented both the simplified version described above and the full Berger and Oliger algorithm. The results are indistinguishable.

## 4. Results

Earlier sections have been terse, but the explanation of our results will, inevitably, take more space. First we explain our modus operandi. Next we define the concept of universality and explain the difference between subcritical and supercritical solutions. We first examine solutions on the axis $r = 0$, because this is relatively simple to visualize. We then extend our discussion to the entire $uv$-plane. Finally we discuss the question of what a distant observer might see, which involves examining the horizon structure (or lack of it).

We have constructed two computer codes to solve the characteristic initial value problem described in section 2. They differ slightly in the treatment of the boundary and initial conditions, but more importantly they use different versions of the Berger-Oliger algorithm as described in section 3, and have different criteria for determining the local truncation error. Within the codes there is a key which mediates the matter-geometry coupling. When this key is switched off the code evolves a test scalar field on Minkowski spacetime, for which the exact solution is known. Both codes reproduce this. With the key switched on we then evolved separately the spacetime using the same initial data. Here the codes produce slightly different answers, as expected because the error criteria are different. However the relative errors are of the order of one part in $10^8$. For comparison the coarsest grid size is $h = 10^{-3}$, and for smooth solutions we expect a local truncation error of $O(h^3)$ and a global error of $O(h^2)$. After convincing ourselves that the codes were solving identical problems we then tried separately a variety of choices of initial data.

Figure 3 shows the evolution of the scalar field $s$ as a function of null coordinates $u$ and $v$ in Minkowski spacetime. As expected the pulse travels in to the axis and bounces out to (future null) infinity leaving a zero field for late $u$. Clearly the shape of the pulse depends on initial data. When the matter-geometry coupling is switched on there is a non-zero "ringing" solution at late $u$. As was first noticed by Choptuik(1992) this ringing solution appears, after the first few oscillations, to be independent of the initial data. We too observe this phenomenon. Features of the solution which are independent of the choice of initial data will be called *universal.*

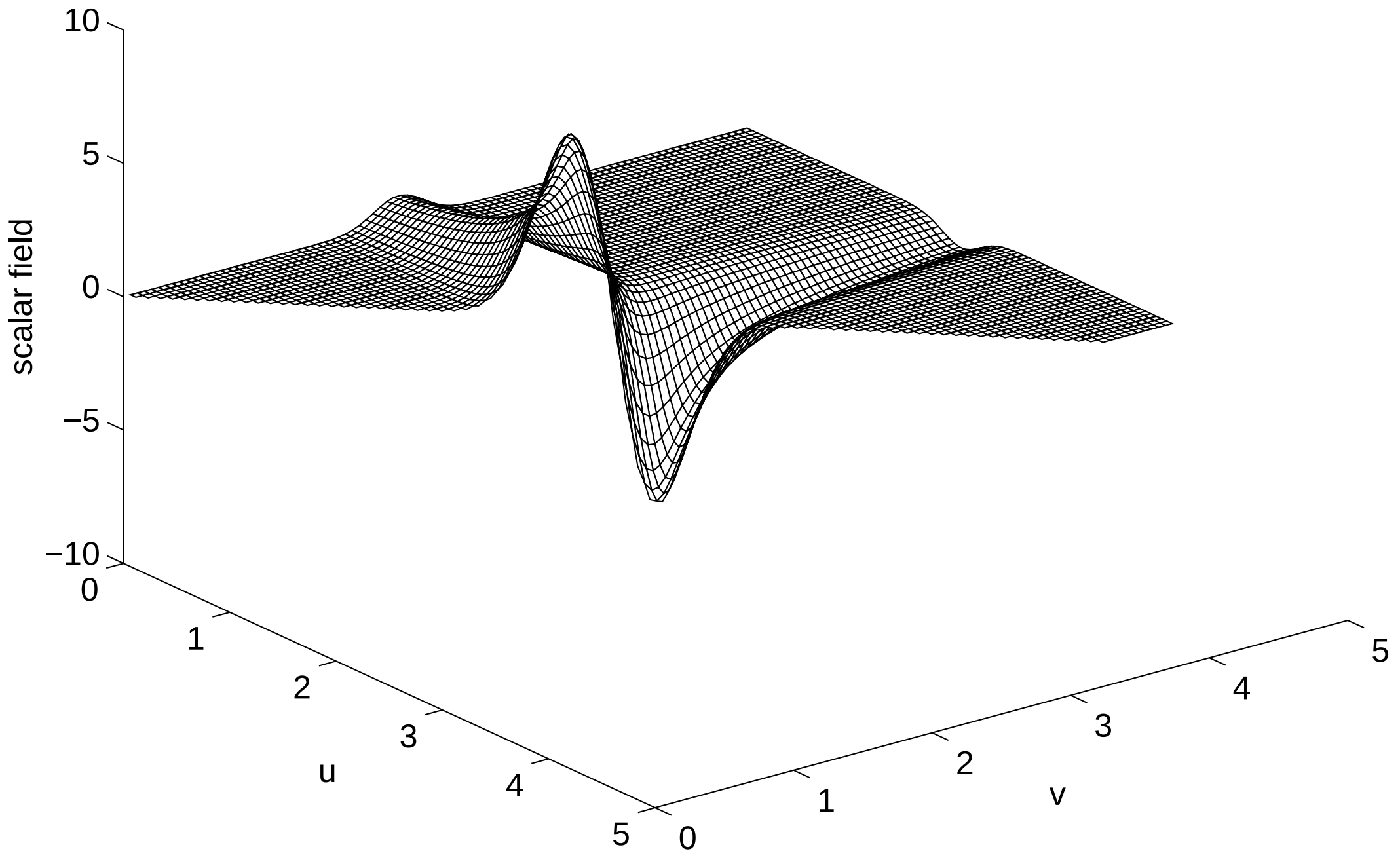

**Fig. 3** The evolution of a test scalar field in Minkowski spacetime as a function of null coordinates $u$ and $v$. The boundaries of the triangular mesh are: (left background) the initial surface $u = 0$, (right background) future null infinity and (foreground) the axis $u = v$. An incoming pulse bounces off the axis and travels to future null infinity leaving an empty spacetime.

As a first example of universality we consider the formation of black holes. We suppose the initial data to depend on a parameter $p$. For small $p$ we have near trivial data and we obtain solutions very like figure 3. For large values of $p$ our evolutions break down because the solutions became singular. One of the variables evolved is $g = \partial r/\partial v$ and the vanishing of $g$ indicates an apparent horizon. For many solutions which become singular say at $(u_s, v_s)$, we find a line of points $(u_h, v_h)$ at which $g = 0$ with $u_h < u_s$. We interpret this as the formation of an apparent horizon surrounding the singular region. We say that a black hole has formed. By performing a sequence of evolutions with different choices for $p$ we are able to locate, almost to machine precision (one part in $10^{16}$), a *critical parameter* $p^*$, which separates black hole formation from field dispersal. Solutions with $p > p^*$ are *supercritical*, those with $p < p^*$ are *subcritical.* For nearly critical solutions it is a non-trivial task deciding their nature. Any solution which remained regular and finally reduced to flat spacetime was deemed to be subcritical. However in some supercritical cases the formation of the apparent horizon and singularity did not appear at coarse grid levels; the coarse spacing failed to resolve the true nature of the solution, which was only discernible from fine grid calculations which occur later in the evolution (step 6 of section 3). Very close to the critical case there are parameters for which we cannot decide what happens. The computers run out of memory before we can ascertain the formation of an apparent horizon or dispersal of the field to infinity. This means that we can only determine the value of $p^*$ to about one part in $10^{15}$. For the supercritical solutions we can compute the Hawking mass given by equation (2.7). We choose a fixed value $v_0$ of $v$ and determine the point $(u_0, v_0)$ where the apparent horizon intersects $v = v_0$. The black hole mass is defined (see below) to be $m_{BH} = m(u_0, v_0)$, and for each one-parameter family of evolutions, it is a function of $p$. Figure 4 shows $\log m_{BH}$ as a function of $\log(p - p^*)$. (NB: throughout this paper all logarithms have base 10 rather than $e$.) There is strong numerical evidence that

$$m_{BH} = K(p - p^*)^{\gamma}, \tag{4.1}$$

where $\gamma \approx 0.374$. If we choose a different parameter for the same initial data, or repeat the process with a completely different choice of initial data, we again obtain the relation (4.1). The parameter $\gamma$ appears to be the same for all evolution families; it is universal. The coefficient $K$ is different for different choices of initial data. Essentially the same result with the same $\gamma$ was reported by Choptuik(1992, 1994). However our code extends the range of validity to black holes 4 orders of magnitude lighter than those of Choptuik. Because $m_{BH} \to 0$ as $p \to p^*$ our typical choices for $v_0$ correspond to $O(10^5)$ "Schwarzschild radii", and so for near criticality our definition of $m_{BH}$ makes sense.

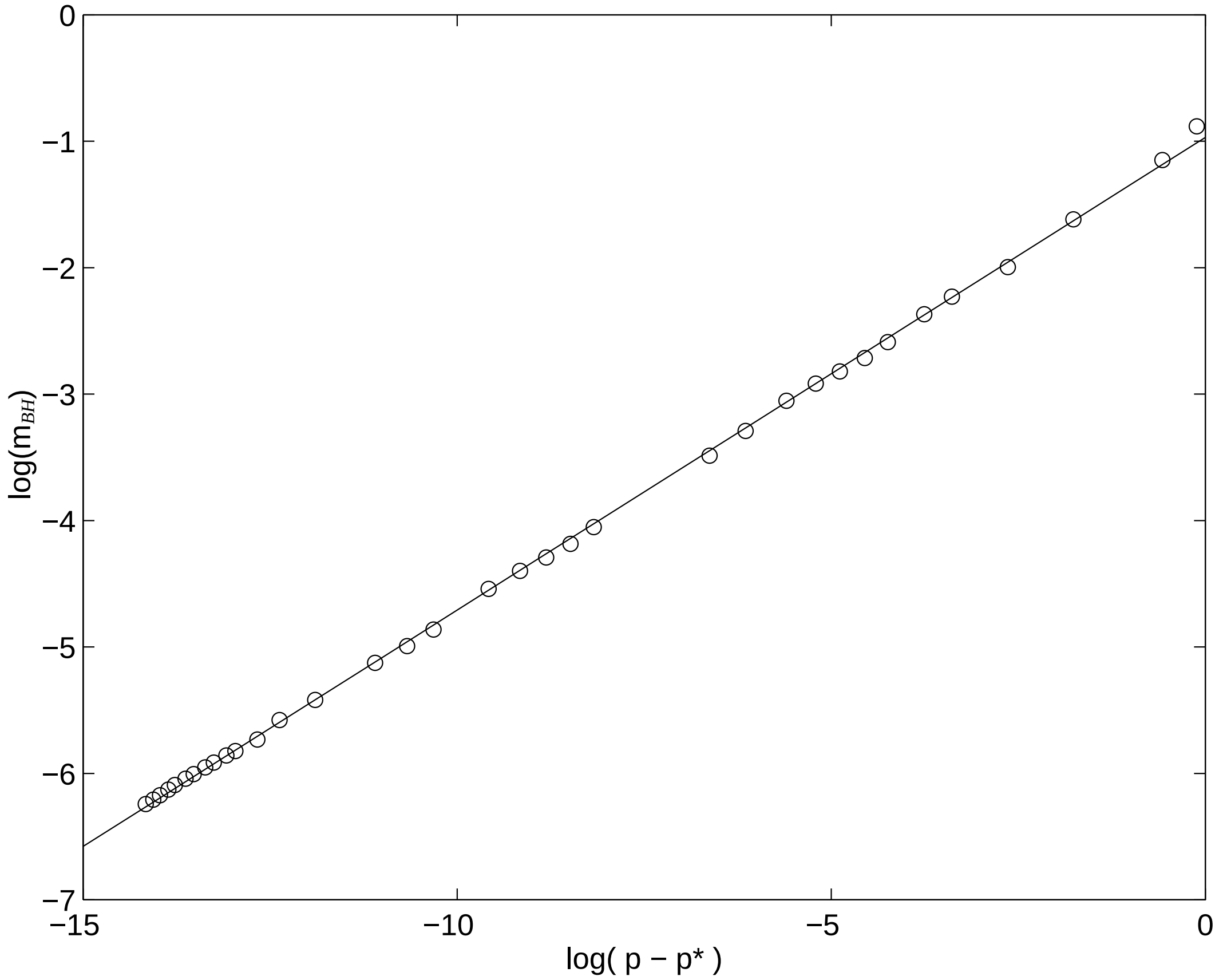

**Fig. 4** The logarithm (base 10) of the black hole mass for supercritical evolutions as a function of the logarithm of $p - p^*$, where $p^*$ is the critical parameter. The best-fit straight line has slope $\gamma = 0.374$.

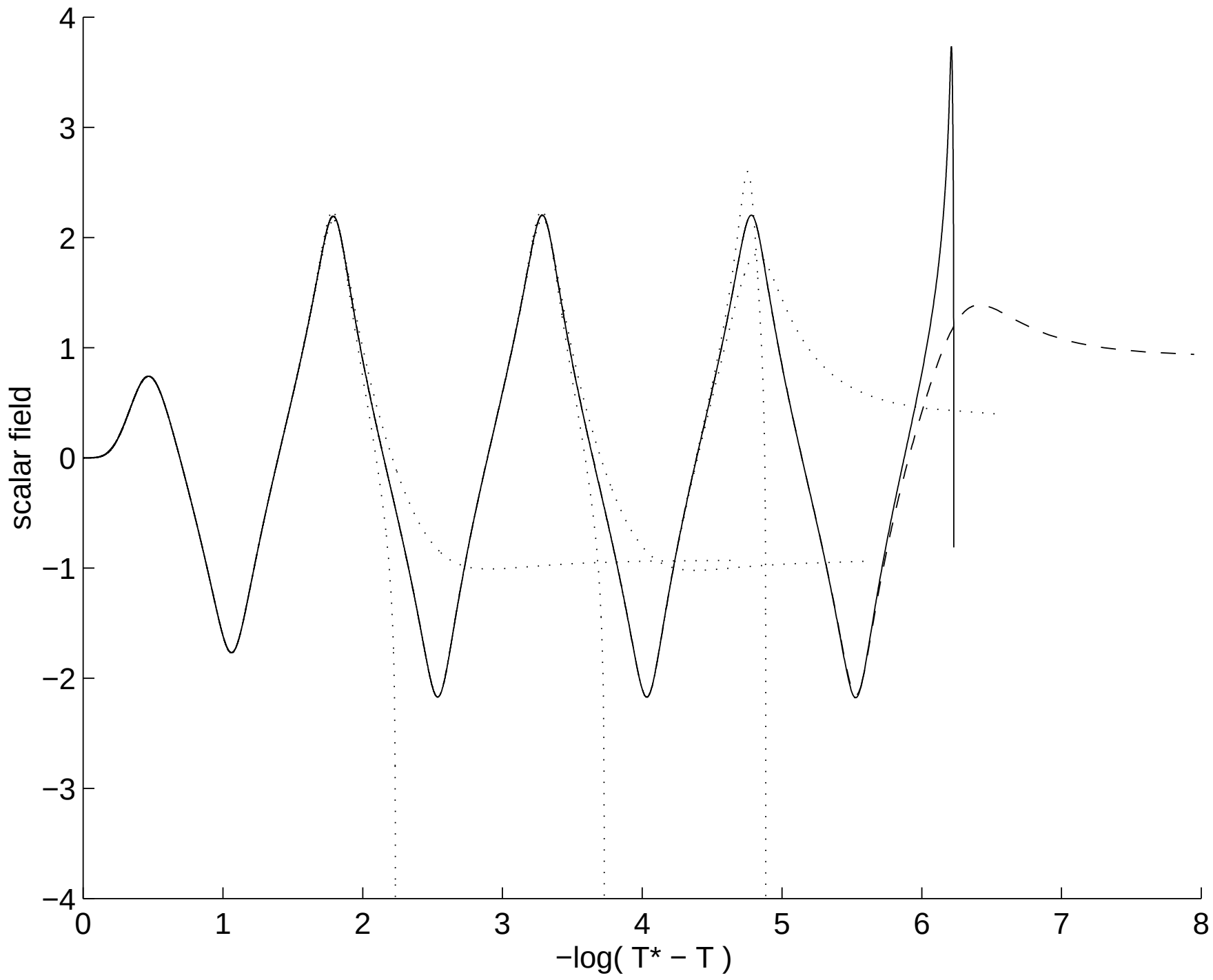

**Fig. 5** The scalar field on axis as a function of $\tau = -\log(T^* - T)$. The solid line represents a marginally supercritical evolution. The first peak and trough show the profile of the initial data. The next three oscillations appear to be universal; they have a period $\Delta\tau = 1.496$. The dashed line represents marginally subcritical data which differs only in the fifth peak. The dotted lines correspond to evolutions which are further from being critical and show fewer oscillations.

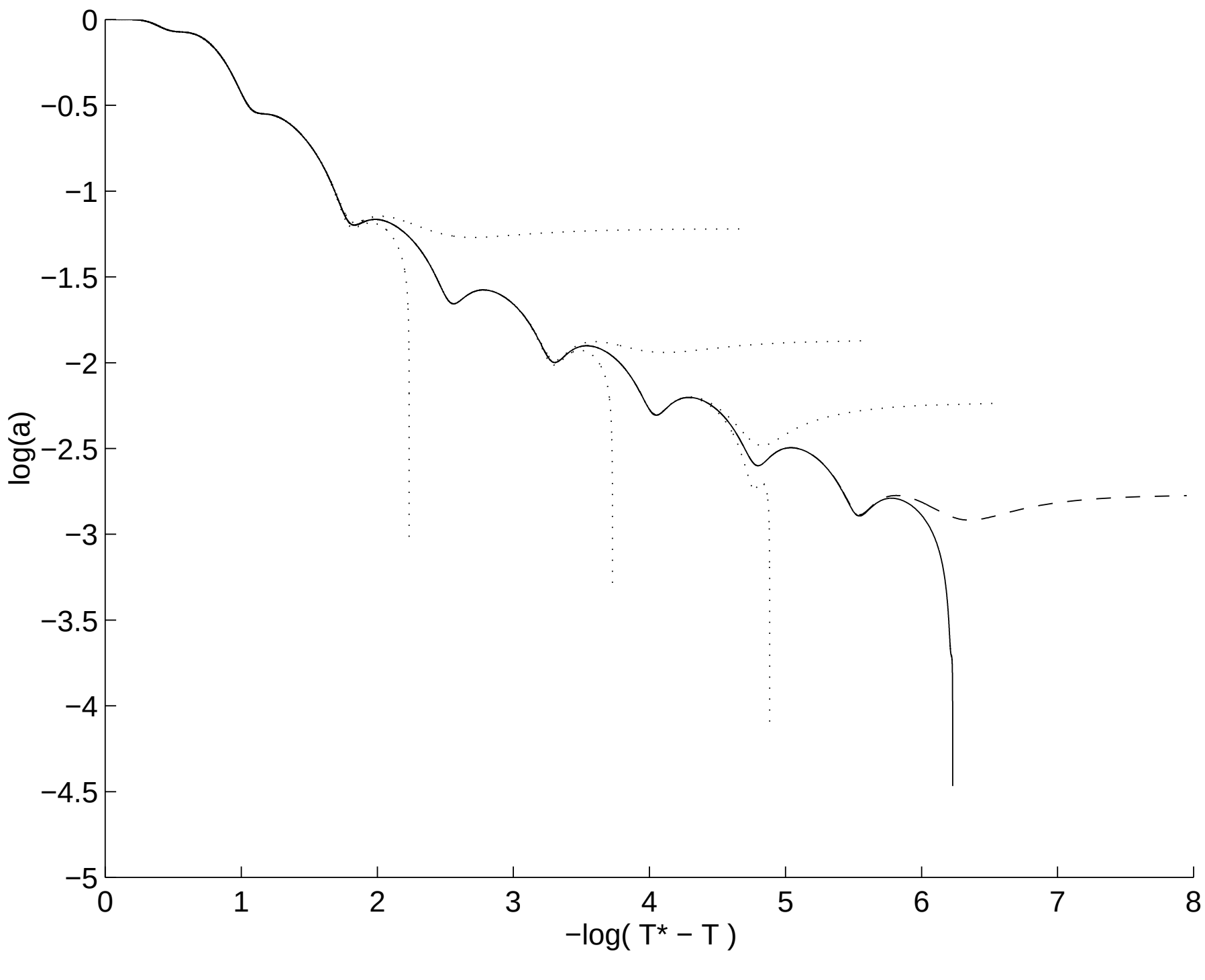

**Fig. 6** The metric function $a$ on axis as a function of $\tau = -\log(T^* - T)$ for the same evolutions as in figure 5. There are twice as many oscillations since $a$ is essentially quadratic in the derivatives of the scalar field. $a \to 0$ at the singularity.

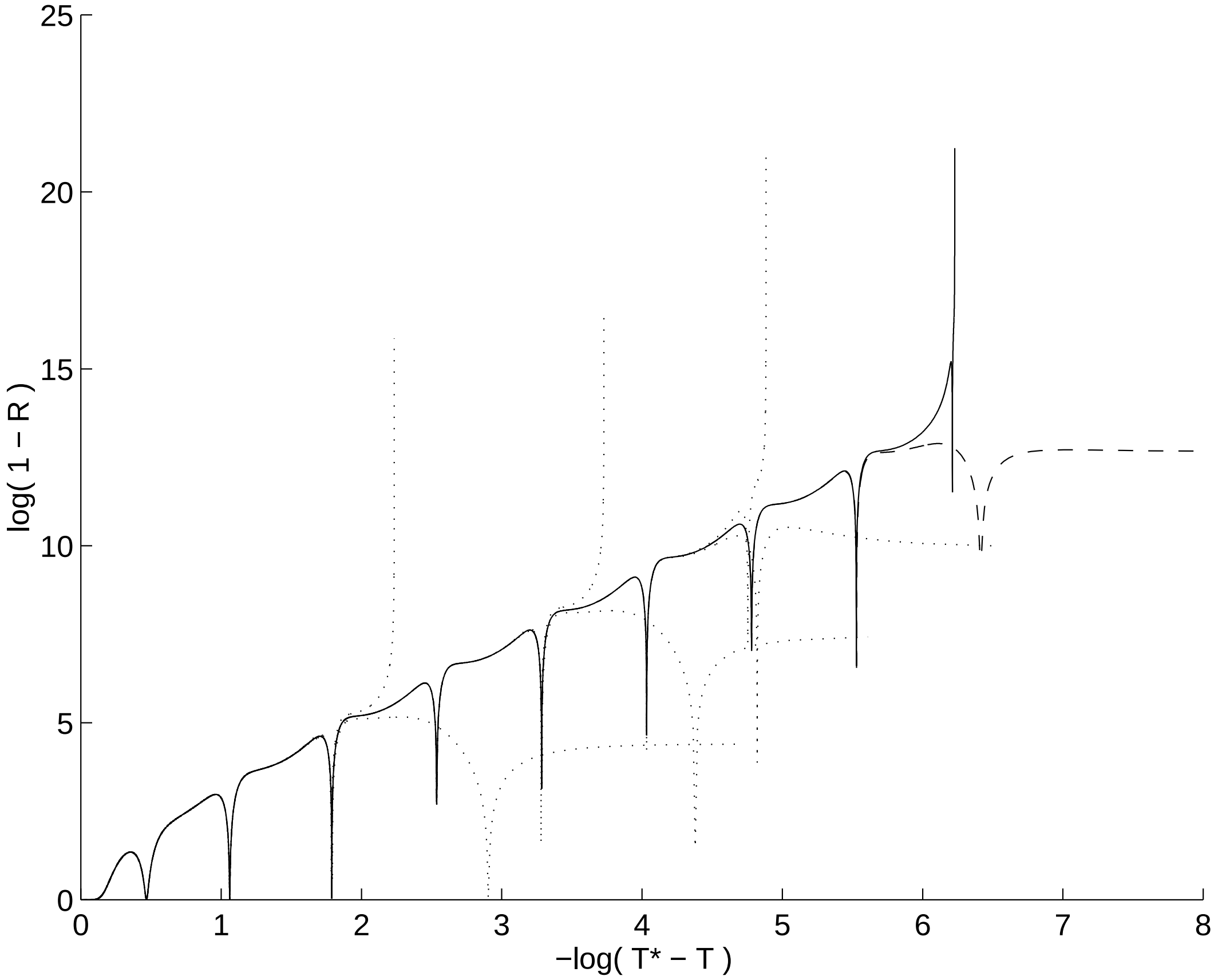

**Fig. 7** The Ricci curvature scalar on axis, plotted as $\log(1-R)$ as a function of $\tau = -\log(T^* - T)$ for the same evolutions as in figure 5. On each oscillation the magnitude of $R$ grows by a factor of approximately 30.

Let us consider next the behaviour of the fields on axis, $u = v$. The function $T$ defined by equation (2.11) measures proper time on axis. Let $T^*$ be the time at which the singularity forms in the exactly critical case. Choptuik(1992, 1994) reported a discrete symmetry, which may be interpreted here as suggesting that the fields on axis should be periodic in the time variable $\tau = -\log(T^* - T)$, with a period $\Delta\tau \approx 1.49$. (We obtain a slightly larger value $\Delta\tau \approx 1.496$. Given the uncertainties in determining $T^*$ there is no inconsistency here.) As seen in $uv$-coordinates the oscillations will appear with the period decreasing by a factor $10^{1.496} \approx 31$ on each oscillation. There is obviously a numerical problem here in that one needs finer and finer grids to resolve the details. In practice our computer resources do not allow us to construct grids with a spacing much less than $10^{-7}$ of our original spacing, and so we might expect to see four but not five oscillations. Figure 5 shows the scalar field $s$ on axis as a function of $\tau$. The solid line represents a slightly supercritical evolution which ends with a black hole. The first peak and trough represent the profile of the initial data. The next three oscillations are periodic and appear to be universal; the same pattern appears whatever the initial data. The dashed line represents a slightly subcritical evolution in which the field eventually dissipates leaving flat spacetime; it differs from the supercritical one only in the fourth oscillation. For these two evolutions the parameter $p$ is very close to the critical one $p^*$. If $p$ is further from $p^*$ the same pattern occurs, but the number of oscillations before black hole formation or dissipation is smaller. Three additional subcritical and three supercritical evolutions are superposed on figure 5 as dotted lines. They show fewer oscillations as $p$ moves further from $p^*$.

This suggests, but does not prove, that if we had the resources to resolve arbitrarily fine grids for near critical evolutions we might expect to see an arbitrary number of oscillations.

Figure 6 shows the metric function $a(u, u)$ as a function of $\tau$ on axis for the same set of evolutions. (There are twice as many oscillations in this and the next picture because the functions plotted are essentially quadratic in the derivatives of the scalar field.) For the supercritical evolutions $a(u, u) \to 0$ extremely quickly as the singularity forms, and we believe that this is why we cannot resolve more oscillations.

An expression for the Ricci curvature scalar was given by equation (2.8) and on axis it can be written as

$$R(u, u) = -32\pi G \frac{d^2\Psi}{dT^2}. \tag{4.2}$$

It is obvious from this form that on each oscillation of the scalar field $R$ will increase by a factor $31^2$, and were there to be an arbitrary number of oscillations then $R$ would grow without bound. This is illustrated in figure 7 where we have plotted $\log(1 - R)$ as a function of $\tau$ for the same evolutions as the two previous figures. (Note that for the subcritical cases the evolution shown here ceases at $T = T^*$, which is before the field disperses to infinity.)

A measure of the energy density of the field is the quantity $\rho$ defined by equation (2.9). On axis this is $-R/(32\pi G)$ and so we deduce that the field energy density becomes large, and possibly unbounded.

Next we consider the fields off-axis. Because many readers are not used to null coordinates we use space and time coordinates. We take $\tau = -\log(T^* - T)$ defined on axis as our time coordinate, and in analogy with the Schwarzschild case we introduce a radial coordinate $r^* = (v - u)/2$. Since for near critical evolutions $r^*$ ranges from 0 to $10^6 m_{BH}$, figures 8–11 use $\operatorname{arcsinh}(200\, r^*)$ as the radial coordinate. In the first two figures time is evolving from the background to the foreground, with the axis as the left border of the mesh and (spatial) infinity as the right border. Figure 8 shows the behaviour of the scalar field in a slightly supercritical evolution. The mesh is a composite picture based on data from all of our grids. In the strong field region the displayed spacing approximately corresponds to points on the second child grid. The details of the formation of the apparent horizon and singularity are hidden in the foreground spike. We shall explore this region in more detail shortly. (See e.g., figures 12–13.) Figure 9 shows the behaviour of the metric function $a(u, v)$. $a = const.$ represents flat spacetime, and so the strong curvature region stays close to the axis. For the next two figures we have, for reasons of clarity, reversed the directions of the time and space axes. Figure 10 shows $\log|1 - R|$ as a function of $\tau$ and $r^*$. The initial data corresponded to flat spacetime and so there is an initially flat region which lasts until the first bounce occurs. Figure 11 shows $\rho$ as a function of $\tau$ and $r^*$ which has similar behaviour.

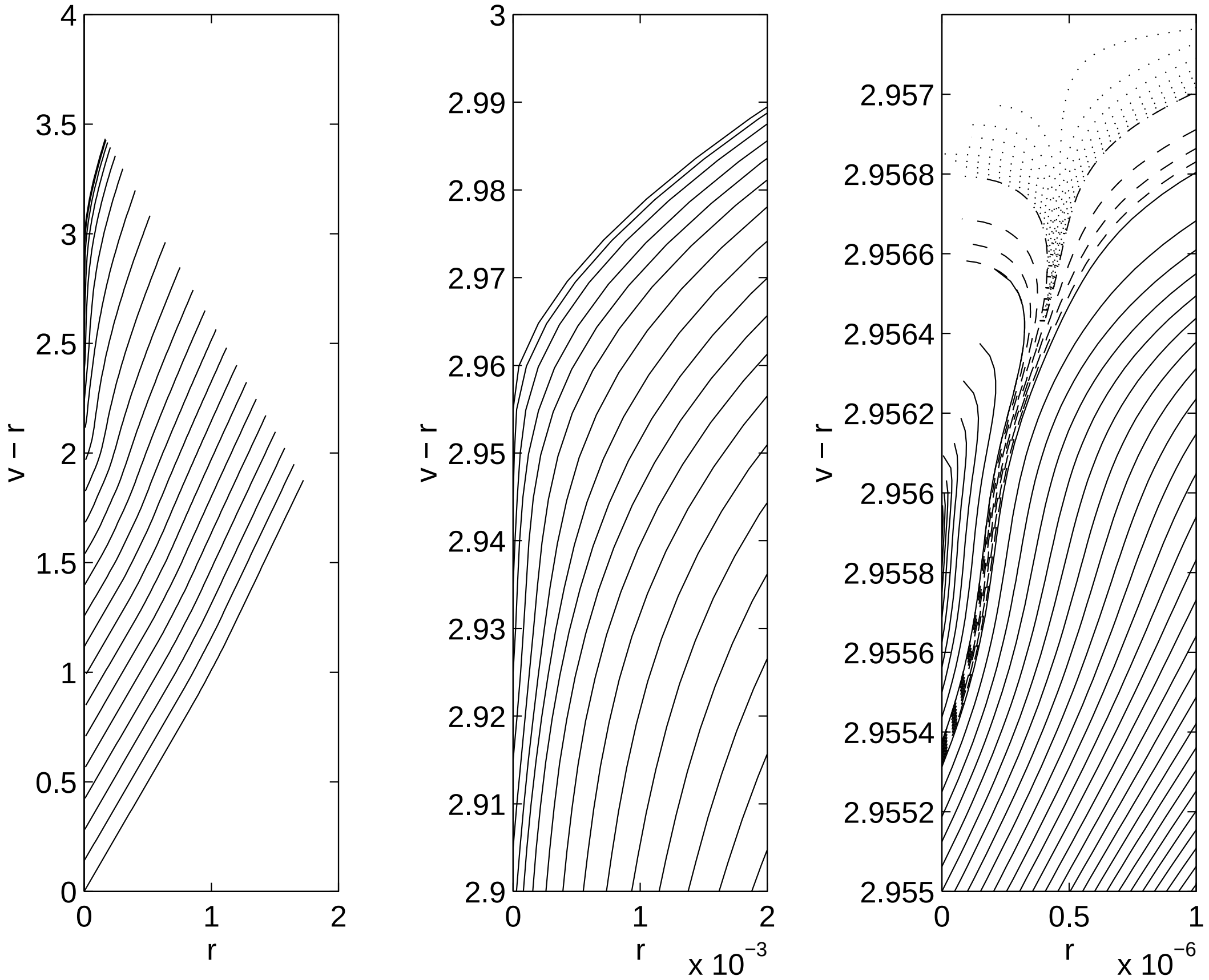

**Fig. 12** The light cone structure for a marginally supercritical solution displayed in ingoing Eddington-Finkelstein coordinates. Each line represents an outgoing cone $u = const.$ and the $u$-spacing is constant. The left diagram is taken from the coarsest grid. There is little to see at this resolution. The middle diagram is an enlargement by a factor of 1000 in each dimension, again taken from the coarsest grid. Finally after an additional expansion by a factor 2000 we see the formation of the horizon and singularity. The dashed lines come from the fourth grid and the dotted lines from the sixth grid. The $u$-spacings are reduced by 4 and 16 respectively.

We have argued that the energy density of the scalar field and the scalar curvature can become arbitrarily large close to the axis. A major question is whether these strong field regions can be seen by a distant observer. Consider an observer who sits at large constant area radius $r$. His proper time is a function of $u$, $T(u,r)$ given by equation (2.10). Of course if an apparent horizon forms, one term in the integrand $(\partial v/\partial u)_r$ becomes infinite and the integral diverges. It takes infinite proper observer time to see the horizon of a black hole. But what happens in a subcritical evolution which is close to critical? A continuity argument might naïvely suggest that it would take a very large (possibly infinite) observer proper time for the strong field region to become visible. Our computations suggest just the opposite.

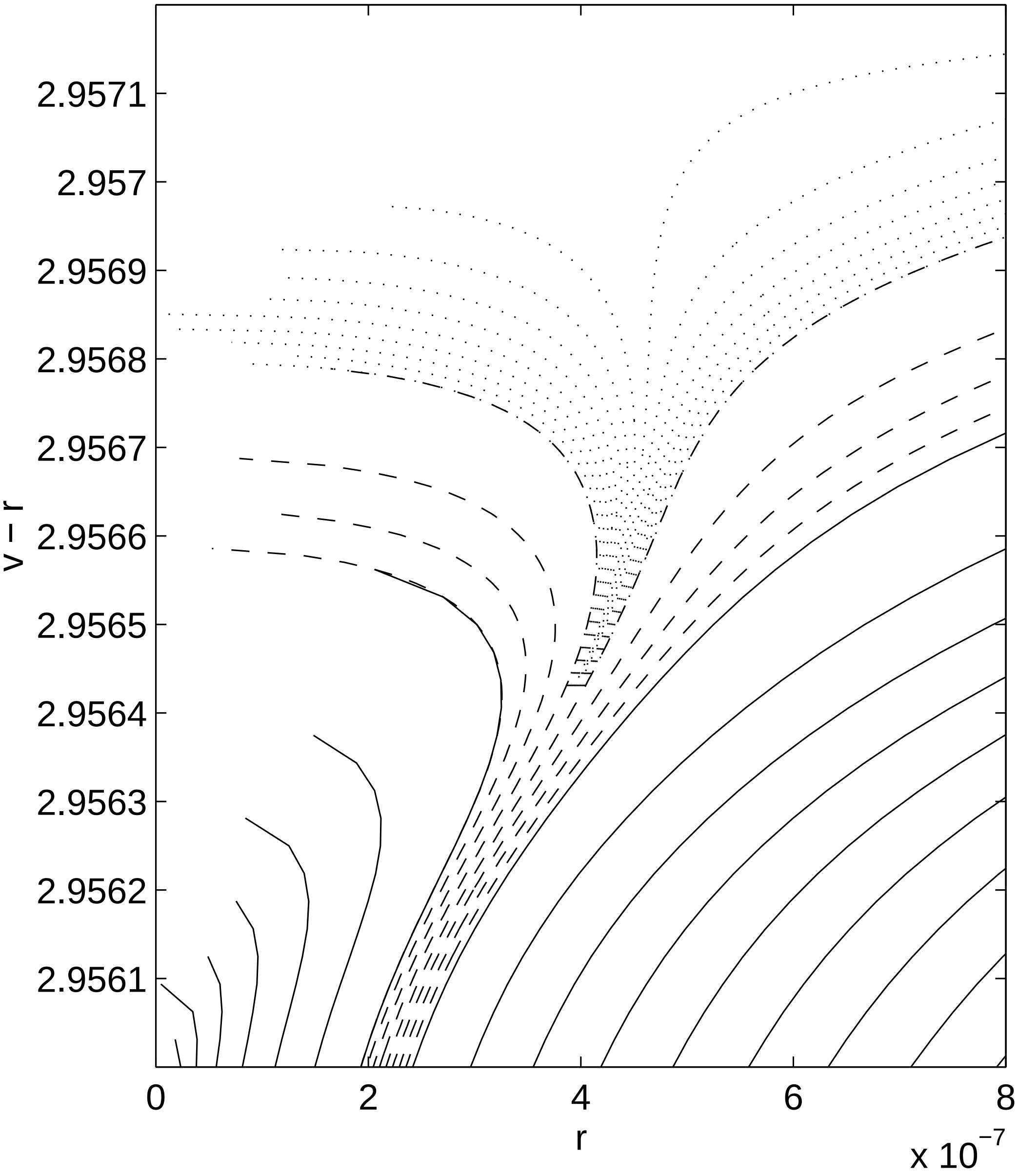

**Fig. 13** This is an enlargement of the right diagram from figure 12. The dashed lines come from the fourth grid and the dotted lines from the sixth grid. The $u$-spacings are reduced by 4 and 16 respectively. Note the speed at which the apparent horizon forms.

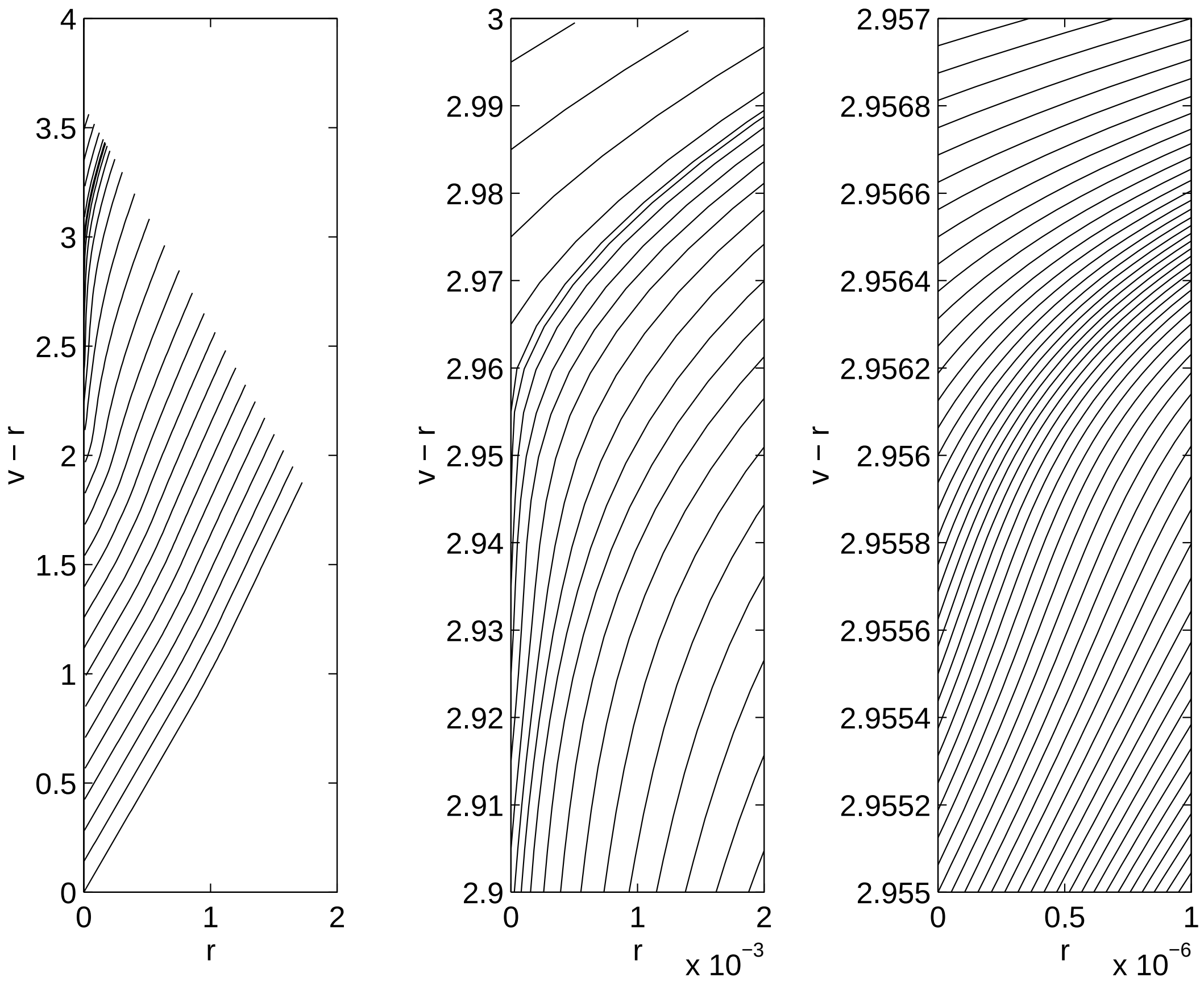

**Fig. 14** The light cone structure for a marginally subcritical solution displayed in ingoing Eddington-Finkelstein coordinates. Each line represents an outgoing cone $u = const.$ and the $u$-spacing is constant. The left diagram is taken from the coarsest grid. There is little to see at this resolution. The middle diagram is an enlargement by a factor of 1000 in each dimension, again taken from the coarsest grid. Note how quickly the cones space out once the strong fields have dispersed. The right diagram is a further expansion by a factor 2000. The parameter $p$ differs from that used in figures 12–13 by $10^{-15}$.

The left diagram in figure 12 shows a neighbourhood of the singularity for a slightly supercritical evolution, in ingoing quasi-Eddington-Finkelstein coordinates. The lines represent surfaces of constant $u$ and the spacing between them is constant. Very little can be seen on this, the coarsest grid. The middle diagram shows part of the diagram enlarged in each dimension by a factor of 1000 again using data from the the coarsest grid. Again there is little to see. The right diagram is an enlargement by a further factor of 2000 and uses data from the third (solid lines) fourth (dashed lines with the $u$-spacing reduced by a factor 4) and sixth (dotted lines with the $u$-spacing reduced by a further factor of 16) subgrid levels. Lines near the right boundary represent outgoing light cones where photons are still able to escape to infinity. As the evolution proceeds these cones fold over and return to the singularity at $r = 0$. An enlargement of part of this diagram is shown as figure 13. Figure 14 covers the same regions as figure 12, but for a slightly subcritical evolution. The difference between the two values of the parameter $p$ is one part in $10^{15}$. The behaviour at late time is quite different. Note, e.g., from the middle diagram how quickly the $u$-spacing increases once the strong field region has dispersed. To emphasize this we have superposed the two evolutions in figure 15. At early times the two evolutions differ by about one part in $10^{15}$ and this near identity is preserved until a time corresponding to the bottom of figure 15 is reached. There is then a clear bifurcation in behaviour as the supercritical evolution (solid lines) forms a horizon. One might have expected the dotted lines of the subcritical evolution to steepen up, mimicking an an apparent horizon, and delaying the progress of an outgoing photon. However the slopes become shallower! Thus one might expect that photons emitted from strong field regions in the subcritical case will reach distant observers moving along paths of constant $r$ at a finite (observer) proper time. This is confirmed by computations of $T(u, r)$ defined by equation (2.10) for a number of subcritical evolutions, which are summarised in figure 16. A value of $r = 0.1$ was chosen and for each evolution the time taken to reach $u = 2.8$, 3.0, 3.2, and 3.4 was computed. (These $u$-values include the strong field region of figures 12–15.) When $p$ is much larger than $p^*$, 0.1 is not a "large" value for $r$ but when $p$ is very close to $p^*$, $r = O(10^5 m_{BH})$ where $m_{BH}$ is the mass of a slightly supercritical black hole. Figure 16 indicates clearly that as $p \to p^*$, $T(u, r) \to$ const. for each relevant value of $u$. Thus a distant observer need wait only a finite proper time in order to see the strong field regions for slightly subcritical evolutions.

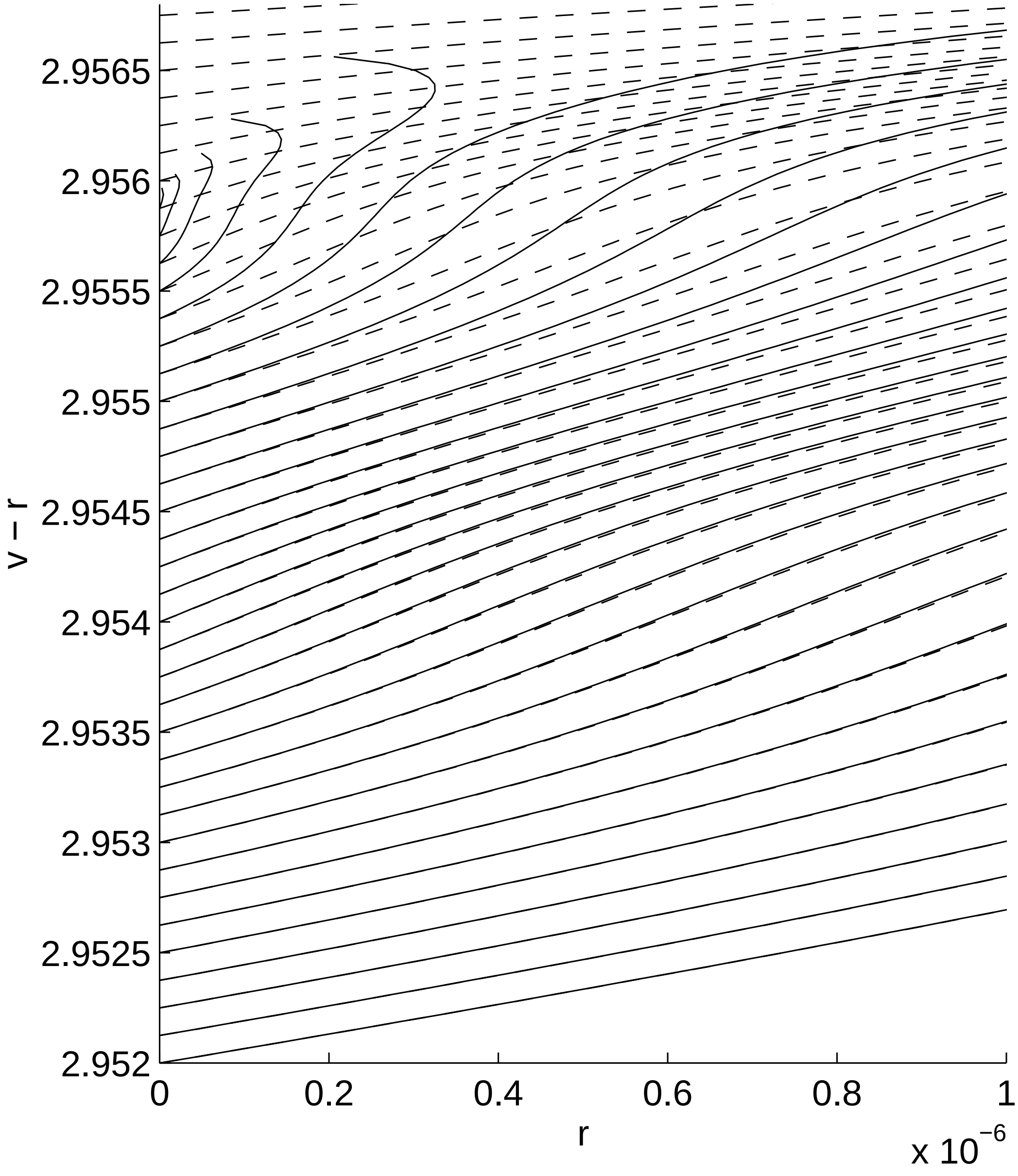

**Fig. 15** This figure superposes the null cone structure of the marginally supercritical solution of figures 12–13 (solid lines) with that of the marginally subcritical one of figure 14 (dotted lines). The initial data differs by one part in $10^{15}$ and for early times (bottom of picture) the results are identical. However at later times, near horizon formation, there is a clear bifurcation between the two cases.

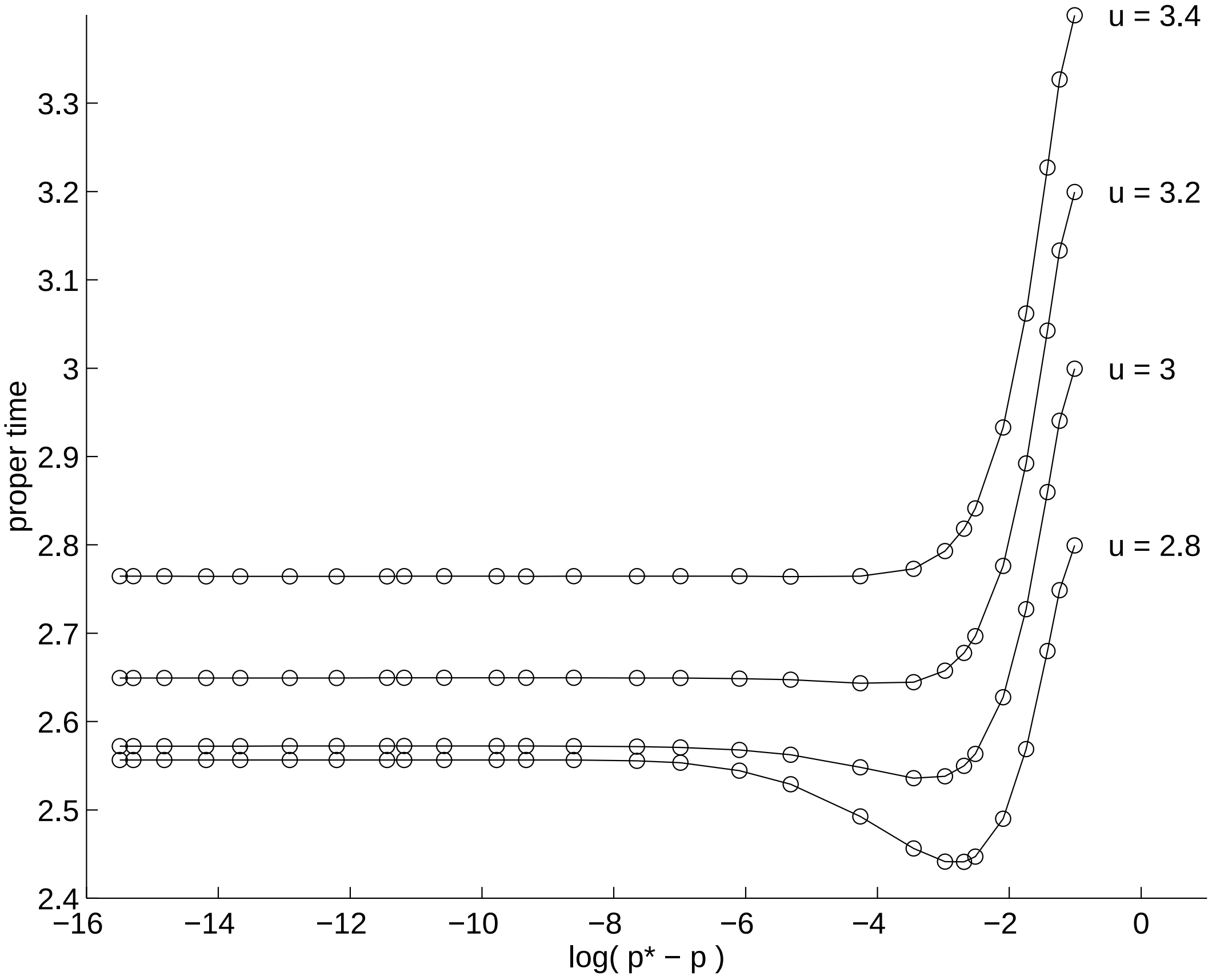

**Fig. 16** For subcritical evolutions with parameter $p$ the proper time taken to reach various values of $u$ (encompassing the strong field region) for an observer at $r = 0.1$, is plotted as a function of $p^* - p$. When $p$ is close to $p^*$, this value of $r$ is $O(10^5 m_{BH})$ where $m_{BH}$ is the black hole mass for slightly supercritical evolutions. Note that as $p \to p^*$ the proper times tend to a finite limit.

## 5. Conclusions

We have implemented a version of the Berger-Oliger algorithm and have used it to study the same scalar field collapse problem as Choptuik(1992, 1994). However our coordinate chart is a double null one which allows evolutions to proceed through the event horizon and up to the singularity in the supercritical case. We have observed the same mass scaling phenomenon, equation (4.1), figure 4, and the echoing property whereby near-critical evolutions display structure on ever finer scales, figures 5–11. Our measured values of universal parameters agree to within expected limits with those of Choptuik, and so we confirm his results.

If the echoing phenomenon really is periodic in $\tau = -\log(T^* - T)$ then it follows from equation (4.2) that on and near the axis the scalar curvature $R$ and scalar field energy density $\rho$ will diverge like $O((T^* - T)^{-2})$ as $T \to T^*$. In the supercritical case this region will be surrounded by an event horizon and therefore invisible to distant observers. However the subcritical case is quite different, figures 12–15, and a distant observer at constant area coordinate $r$ will be able to see photons emitted from the strong field region after a finite proper time, figure 16.

One could argue that our physical scheme, equations (1.1–2), is only a toy model and that for more realistic situations the strong field regions would not be visible. Checking this would be more complicated.

Further details of these evolutions and those for more complicated field configurations will be published elsewhere.

## Acknowledgments

We would like to thank James Horne for useful discussions and Stephen Hawking for asking awkward questions and encouraging us to find the answers. R.S.H. was supported by an EPSRC research studentship. All of this work was done on workstations supplied by the EPSRC (formerly SERC) Computational Science Initiative Research Grant GR/H57585.